\documentstyle[epsfig]{article}
\begin{document}

\title{\Large \bf  Notes for a brief history of quantum gravity\\
{\normalsize \rm \em Presented at the 9th Marcel Grossmann 
Meeting in Roma, July 2000}}

\author{
\normalsize\rm \bf Carlo Rovelli\\
\normalsize\em Centre de Physique Th\'eorique, 
CNRS Luminy, 13288 Marseille, France \\
\normalsize\em Physics Department, University of 
Pittsburgh, Pittsburgh, PA 15260, USA.\rm
\\
\normalsize\textsf{rovelli@cpt.univ-mrs.fr }}
\date{\normalsize \today}
\maketitle

\begin{abstract}
\noindent 
I sketch the main lines of development of the research in 
quantum gravity, from the first explorations in the early 
thirties to nowadays.  
\end{abstract}

\section{Introduction}

When John Stachel asked me to prepare a brief history of the research
in quantum gravity for the 9th Marcel Grossmann Meeting, I trembled at
the size of the task, worried of repeating only information already
known to everybody, and feared to displease my colleagues.  John
managed to convince me to try anyway, and here is the result.  I have
much enjoyed spending time in the ``old archives" section of my
library, and I have been surprised by some of the things I have found. 
I am enormously indebted with the many friends that, after the archive
posting of the first draft of this work, have pointed our errors,
omissions and imperfections.  With their invaluable help, this history
is a bit less biased and a bit less incomplete. 

I have focused on quantum gravity in the strict sense: the search for
a theory that could describes the quantum behavior of the full
gravitational field.  Thus, I do not cover important related subjects
such as quantum fields in curved spacetime, applications such as
cosmology related research, work on the structure of quantum
constrained systems, black hole thermodynamics or extensions of
quantum mechanics to general covariant theories.  For lack of energy,
I have also decided not to cover the numerical and lattice-like
approaches to the theory -- which is a serious absence.

I have no ambition of presenting complete references to all the
important works; some of the references are to original works, others
to reviews where reference can be found.  Errors and omissions are
unfortunately unavoidable and I apologize for these.  I have made my
better effort to be balanced, but in a field that has not yet
succeeded in finding consensus, my perspective is obviously
subjective.  Trying to write history in the middle of the developments
is hard.  Time will go along, dust will settle, and it will slowly
become clear if we are right, if some of us are right, or --a
possibility never to disregard-- if we all are wrong.

\section{Three main directions}

An evident peculiarity of the research in quantum gravity is that all
along its development it can be separated into three main lines of
research.  The relative weight of these lines has changed, there have
been important intersections and connections between the three, and
there has been research that does not fit into any of the three lines. 
Nevertheless, the three lines have maintained a distinct individuality
across 70 years of research.  The three main lines are often denoted
``covariant", ``canonical", and ``sum over histories", even if these
names can be misleading and are often confused.  They cannot be
characterized by a precise definition, but within each line there is a
remarkable methodological unity, and a remarkable consistency in the
logic of the development of the research.

\begin{description}
    \item[The covariant line of research] is the attempt to build the
    theory as a quantum field theory of the fluctuations of the metric
    over a flat Minkowski space, or some other background metric
    space.  The program was started by Rosenfeld, Fierz and Pauli in
    the thirties.  The Feynman rules of general relativity (GR, from
    now on) were laboriously found by DeWitt and Feynman in the
    sixties.  t'Hooft and Veltman, Deser and Van Nieuwenhuizen, and
    others, found firm evidence of non-renormalizability at the
    beginning of the seventies.  Then, a search for an extension of GR
    giving a renormalizable or finite perturbation expansion started. 
    Through high derivative theory and supergravity, the search
    converged successfully to string theory in the late eighties.
    
    \item[The canonical line of research] is the attempt to construct
    a quantum theory in which the Hilbert space carries a
    representation of the operators corresponding to the full metric,
    or some functions of the metric, without background metric to be
    fixed.  The program was set by Bergmann and Dirac in the fifties. 
    Unraveling the canonical structure of GR turned out to be
    laborious.  Bergmann and his group, Dirac, Peres, Arnowit Deser and
    Misner completed the task in the late fifties and early sixties. 
    The formal equations of the quantum theory were then written down
    by Wheeler and DeWitt in the middle sixties, but turned out to be
    too ill-defined.  A well defined version of the same equations was
    successfully found only in the late eighties, with loop quantum
    gravity.

    \item[The sum over histories line of research] is the attempt to
    use some version of Feynman's functional integral quantization to
    define the theory.  Hawking's Euclidean quantum gravity,
    introduced in the seventies, most of the the discrete
    (lattice-like, posets \ldots) approaches and the spin foam models,
    recently introduced, belong to this line.

\item[Others.]  There are of course other ideas that have been
explored: \begin{itemize} \item Twistor theory has been more fruitful
on the mathematical side than on the strictly physical side, but it is
still actively developing.  \item Noncommutative geometry has been
proposed as a key mathematical tool for describing Planck scale
geometry, and has recently obtained very surprising results,
particularly with the work of Connes and collaborators.  \item
Finkelstein, Sorkin, and others, pursue courageous and intriguing
independent paths.  \item Penrose idea of a gravity induced quantum
state reduction have recently found new life with the perspective of a
possible experimental test.  \item \ldots
\end{itemize} 
So far, however, none of these alternatives has been developed into a
large scale research program.
\end{description}

\section{Five periods}

Historically, the evolution of the research in quantum gravity can
roughly be divided in five periods. 
\begin{description}

    \item[The Prehistory: 1930-1959.]  The basic ideas of all 
    three lines of research appear very early, already in the 
    thirties.  By the end of the fifties the three research 
    programs are clearly formulated.
    
    \item[The Classical Age: 1960-1969.] The sixties see the strong
    development of two of the three programs, the covariant and the
    canonical.  At the end of the decade, the two programs have both
    achieved the basic construction of their theory: the Feynman rules
    for the gravitational field on one side and the Wheeler-DeWitt
    equation on the other.  To get to these beautiful results, an 
    impressive amount of technical labour and ingenuity has proven
    necessary.  The sixties close --as they did in many other
    regards-- with the promise of a shining new world.
        
    \item[The Middle Ages: 1970-1983.] The seventies soon 
    disappoint the hopes of the sixties.  It becomes increasingly
    clear that the Wheeler-DeWitt equation is too ill defined for
    genuinely field theoretical calculations.  And evidence for the
    non-renormalizability of GR piles up.  Both lines of attach have
    found their stumbling block.
    
    In 1974, Steven Hawking derives black hole radiation.  Trying to
    deal with the Wheeler-DeWitt equation, he develops a version of
    the sum over history as a sum over ``Euclidean" (Riemannian)
    geometries.  There is excitement with the idea of the wave
    function of the universe and the approach opens the way for
    thinking and computing topology change.  But for field theoretical
    quantities the euclidean functional integral will prove as weak a
    calculation tool as the Wheeler-DeWitt equation.
    
    On the covariant side, the main reaction to non-renormalizability
    of GR is to modify the theory.  Strong hopes, then disappointed,
    motivate extensive investigations of supergravity and higher
    derivative actions for GR. The landscape of quantum gravity is
    gloomy.
    
    \item[The Renaissance: 1984-1994.]  Light comes back in the 
    middle of the eighties.  In the covariant camp, the various 
    attempts to modify GR to get rid of the infinities merge into 
    string theory.  Perturbative string theory finally delivers 
    on the long search for a computable perturbative theory for 
    quantum gravitational scattering amplitudes.  To be sure, 
    there are prices to pay, such as the wrong dimensionality of 
    spacetime, and the introduction of supersymmetric particles, 
    which, year after year, are expected to be discovered but, so 
    far, are not.  But the result of a finite perturbation 
    expansion, long sought after, is to good to be discarded just 
    because the world insists in looking different from our 
    theories.
    
    Light returns to shine on the canonical side as well.  Twenty 
    years after the Wheeler-DeWitt equation, loop quantum gravity 
    finally provides a version of the theory sufficiently well 
    defined for performing explicit computations.  Here as well, 
    we are far from a complete and realistic theory, and 
    scattering amplitudes, for the moment, can't be computed at 
    all, but the excitement for having a rigorously defined, 
    nonperturbative, general covariant and background independent 
    quantum field theory, in which physical expectation values 
    can be computed, is strong. 
    
     \item[Nowadays: 1995-2000.]  Both string theory and loop 
     quantum gravity grow strongly for a decade, until, in the 
     middle of the nineties, they begin to deliver physical 
     results.  The Bekenstein-Hawking black hole entropy formula 
     is derived within both approaches, virtually simultaneously.  
     Loop quantum gravity leads to the computation of the first 
     Planck scale quantitative physical predictions: the spectra 
     of the eigenvalues of area and volume.
    
    The sum over histories tradition, in the meanwhile, is not 
    dead.  In spite of the difficulties of the euclidean 
    integral, it remains as a reference idea, and guides the 
    development of several lines of research, from the discrete 
    lattice-like approaches, to the ``state sum" formulation of 
    topological theories.  Eventually, the last motivate the spin 
    foam formulation, a translation of loop quantum gravity into 
    a Feynman sum over histories form.
    
     \newpage
  \addtolength{\topmargin}{-2.5cm}
  \thispagestyle{empty}
   \centerline{\psfig{figure=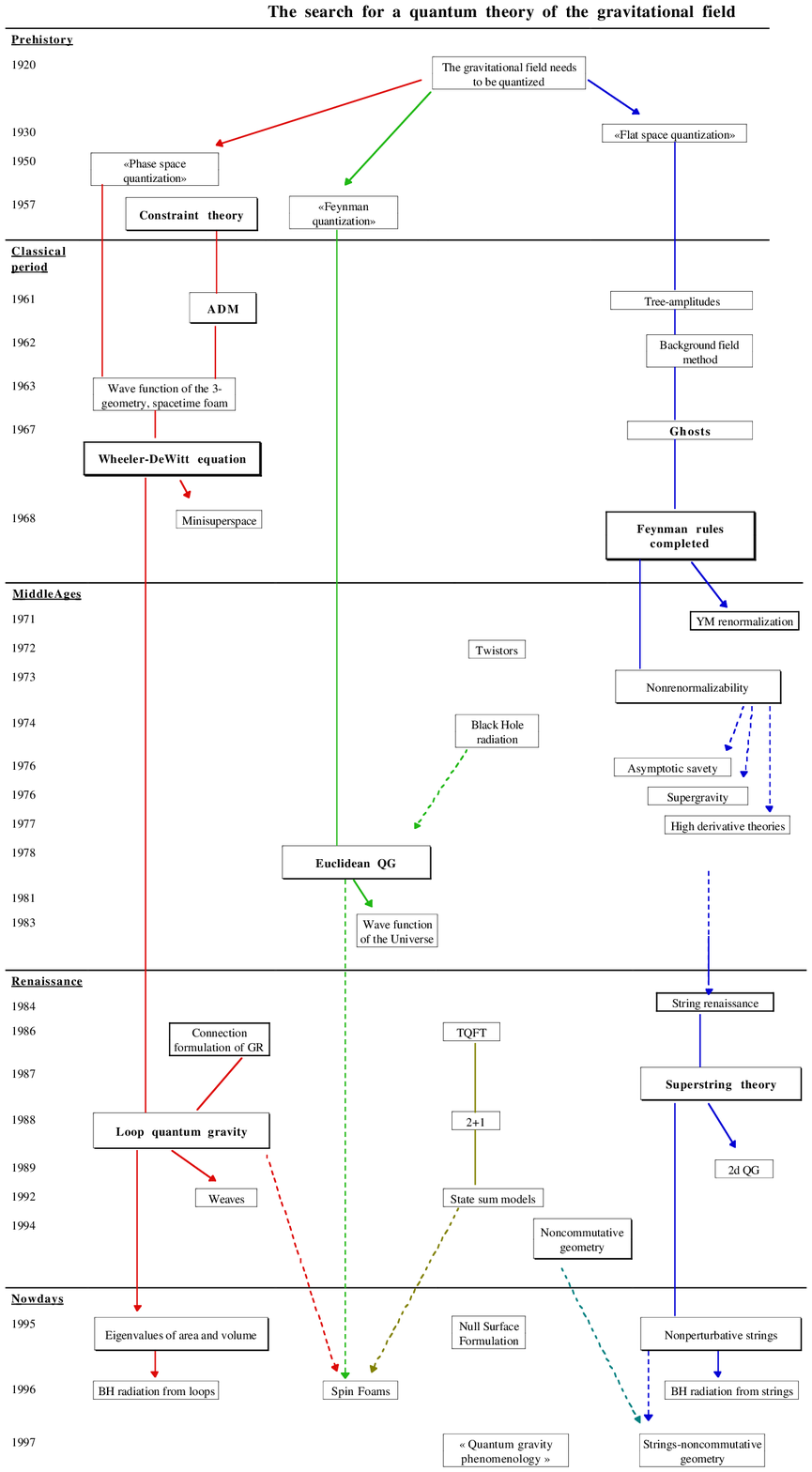}}
  \newpage
  \addtolength{\topmargin}{2.5cm}
  
    Other ideas develop in the meanwhile, most notably noncommutative
    geometry, which finds intriguing points of contact with string
    theory towards the end of the decade.

    The century closes with two well developed contenders for a
    quantum theory of gravity: string theory and loop quantum gravity,
    as well as a set of intriguing novel new ideas that go from
    noncommutative geometry to the null surfaces formulation of GR, to
    the attempt to merge strings and loops.  And even on a very
    optimistic note: the birth of a new line of research, the
    self-styled ``quantum gravity phenomenology" which investigates
    the possibility --perhaps not so far fetched-- that Planck scale
    type measurements might be within reach.  And thus that we could
    finally perhaps know which of the theoretical hypotheses, if any,
    make sense.

\end{description}

Let me now describe the various periods and their main steps
in more detail.

\section{The Prehistory: 1930-1957}

General relativity was found in 1915.  Quantum mechanics in 1926.  A
few years later, around 1930, Born, Jordan and Dirac are already
capable of formalizing the quantum properties of the electromagnetic
field.  How long did it take to realize that the gravitational field
should --most presumably-- behave quantum mechanically as well? 
Almost no time: already in 1916 Einstein points out that quantum
effects must lead to modifications in the theory of general relativity
\cite{einstein}.  In 1927 Oskar Klein suggests that quantum gravity
should ultimately modify the concepts of space and time \cite{klein}. 
In the early thirties Rosenfeld \cite{rosenfeld} writes the first
technical papers on quantum gravity, applying Pauli method for the
quantization of fields with gauge groups to the linearized Einstein
field equations.  The relation with a linear spin-two quantum field is
soon unraveled in the works of Fierz and Pauli \cite{fierz} and the
spin-two quantum of the gravitational field, presumably first named
``graviton'' in a 1934 paper by Blokhintsev and Gal'perin \cite{bg},
is already a familiar notion in the thirties.  Bohr considers the idea
of identifying the neutrino and the graviton.  In 1938, Heisenberg
\cite{heisenberg} points out that the fact that the gravitational
coupling constant is dimensional is likely to cause problems with the
quantum theory of the gravitational field.

The history of these early explorations of the quantum properties of
spacetime has been recently reconstructed in a wonderful and
fascinating paper by John Stachel \cite{stachel}.  In particular, John
describes in his paper the extensive, but largely neglected, work
conducted in the mid thirties by a Russian physicist, Matvei Petrovich
Bronstein.  Bronstein, (who was nephew of Leon Trozky) re-derives the
Rosenfeld-Pauli quantization of the linear theory, but realizes that
the unique features of gravitation require a special treatment, when
the full nonlinear theory is taken into account.  He realizes that
field quantization techniques must be generalized in such a way as to
be applicable in the absence of a background geometry.  In particular,
he realizes that the limitation posed by general relativity on the
mass density radically distinguishes the theory from quantum
electrodynamics and would ultimate lead to the need to ``reject
Riemannian geometry" and perhaps also to ```reject our ordinary
concepts of space and time" \cite{bronstein}.  For a detailed
discussion of Bronstein early work in quantum gravity see ref
\cite{gore}.  The reason Bronstein has remained unknown for so long
has partly to do with the fact that he was executed by the Soviet
State Security Agency (the NKVD) at the age of 32.  In Russia, even
today Bronstein is remembered by some as ``smarter than Landau". 
References and many details on these pioneering times are in this
fascinating paper by John, which I strongly recommend to the reader. 
Here, I pick up the historical evolution from after World War II. In
particular, I start from 1949, a key year for the history of quantum
gravity.

\vskip2mm{\bf 1949}\nopagebreak

Peter Bergmann starts its program of phase space quantization of 
non linear field theories \cite{bergmann49}.  He soon realizes 
that physical quantum observables must correspond to coordinate 
independent quantities only \cite{bergmanOB}.  The search for 
these gauge independent observables is started in the group that 
forms around Bergmann, at Brooklyn Polytechnic and then in 
Syracuse.  For instance, Ted Newman develops a perturbation 
approach for finding gauge invariant observables order by order 
\cite{ted}.  The group studies the problems raised by systems 
with constraints and reaches a remarkable clarity, unfortunately 
often forgotten later on, on the problem of what are the 
observables in general relativity.  The canonical approach to 
quantum gravity is born.

Bryce DeWitt completes his thesis.  He applies Schwinger's covariant
quantization to the gravitational field.

Dirac presents his method for treating constrained hamiltonian systems
\cite{dirac50}.

\vskip2mm{\bf 1952}\nopagebreak

Following the pioneering works of Rosenfeld, Fierz and Pauli, 
Gupta \cite{gupta} develops systematically the ``flat space 
quantization" of the gravitational field.  The idea is simply to 
introduce a fictitious ``flat space", that is, Minkowski metric 
$\eta_{\mu\nu}$, and quantize the small fluctuations of the 
metric around Minkowski $h_{\mu\nu}=g_{\mu\nu}-\eta_{\mu\nu}$.  
The covariant approach is fully born.  The first difficulty is 
immediately recognized, in searching the propagator, as coming 
from the fact that the quadratic term of the Lagrangian is 
singular, as for the electromagnetic field, and as a consequence 
of gauge invariance.  Gupta's treatment uses an indefinite norm 
state space as for the electromagnetic field.

\vskip2mm{\bf 1957} \nopagebreak

Charles Misner introduces the ``Feynman quantization of general
relativity" \cite{misner}.  He quotes John Wheeler for suggesting the
expression
\begin{equation}
    \int exp\{(i/\hbar)({\rm Einstein\ action)}\}\ d({\rm field\ 
    histories)}),
    \label{eq:functional}
\end{equation}
and studies how to have a well defined version of this idea.  Misner's
paper \cite{misner} is very remarkable in many respects.  It explains
with complete clarity notions such as why the quantum hamiltonian must
be zero, why the individual spacetime points are not defined in the
quantum theory and the need of dealing with gauge invariance in the
integral.  Even more remarkably, the paper opens with a discussion of
the possible directions for quantizing gravity, and lists the three
lines of directions, covariant, canonical, and sum over histories,
describing them almost precisely with the same words we would
today!\footnote{To be sure, Misner lists a 4th approach as well, based
on the Schwinger equations for the variation of the propagator, but
notices that ``this method has not been applied independently to
general relativity", a situation that, as far as I know, has not
changed since.}

\vskip.5cm

At the end of the fifties, all the basic ideas and the research
programs are clear.  It is only a matter of implementing them, and
seeing if they work.  The implementation, however, turns out to be a
rather herculean task, that requires the ingenuity of people of the
caliber of Feynman and DeWitt on the covariant side, and of Dirac and
DeWitt, on the canonical side.

\section{The Classical Age: 1958-1969}

\hspace{.6cm}{\bf 1950} \nopagebreak

The Bergmann group, and Dirac \cite{dirac58}, work out the general
hamiltonian theory of constrained system.  For a historical
reconstruction of this achievement, see \cite{bergmanRE}.  At the
beginning, Dirac and the Bergmann group work independently.  The
present double classification in primary and secondary constraints and
in first and second class constraints, still reflects this original
separation.

\vskip2mm{\bf 1959} \nopagebreak

By 1959, Dirac has completely unraveled the canonical structure of GR.

\vskip2mm{\bf 1961} \nopagebreak

Arnowit, Deser and Misner complete what we now call the ADM 
formulation of GR, namely its hamiltonian version in appropriate 
variables, which greatly simplify the hamiltonian formulation and 
make its geometrical reading transparent \cite{ADM}.

In relation to the quantization, Arnowit, Deser and Misner 
present an influential argument for the finiteness of the self 
energy of a point particle in classical GR and use it to argue 
that nonperturbative quantum gravity should be finite.

\vskip2mm{\bf 1962} \nopagebreak

Feynman attacks the task of computing transition amplitudes in quantum
gravity.  He shows that tree-amplitudes lead to the physics one
expects from the classical theory \cite{feynman62}.

DeWitt starts developing his background field methods for the
computation of perturbative transition amplitudes \cite{dewitt62}.

Bergmann and Komar clarify what one should expect from a Hilbert 
space formulation of GR \cite{bergmankomar}.

Following the ADM methods, Peres writes the Hamilton-Jacobi
formulation of GR \cite{peres}
$$
G^{2}(q_{ab}q_{cd}-{1\over 2}q_{ac}q_{bd})\ {\delta S(q)\over \delta 
q_{ac}}{\delta S(q)\over \delta q_{bd}} + \det q\,R[q]=0,
$$
which will lead to the Wheeler-DeWitt equation.  $q_{ab}$
is the ADM 3-metric and $G$ the Newton constant.

\vskip2mm{\bf 1963} \nopagebreak

John Wheeler realizes that the quantum fluctuations of the 
gravitational field must be short scale fluctuations of the 
geometry and introduces the physical idea of spacetime foam 
\cite{wheeler63}.  Wheeler's Les Houches lecture note are 
remarkable in many respects, and are the source of many of the 
ideas still current in the field.  Just to mention two others: 
``Problem 56" suggests that gravity in 2+1 dimensions may not be 
so trivial after all, and indicates it may be an interesting 
model to explore.  ``Problem 57" suggests to study quantum 
gravity by means of a Feynman integral over a spacetime lattice.

\vskip2mm{\bf 1964} \nopagebreak

Penrose introduces the idea of spin networks, and of a discrete
structure of space controlled by $SU(2)$ representation theory.  The
construction exists only in the form of a handwritten manuscript.  It
gets published only in 1971 \cite{Penrose2}.  The idea will
surprisingly re-emerge 25 years later, when spin networks will be
found to label the states of loop quantum gravity \cite{spinnetworks}.

\vskip2mm{\bf 1964} \nopagebreak

Beginning to study loop corrections to GR amplitudes, Feynman observed
that unitarity was lost for naive diagrammatic rules.  DeWitt
\cite{dewitt64} develops the combinatorial means to correct the
quantisation (requiring independence of diagrams from the longitudinal
parts of propagators).  These correction terms can be put in the form
the form of loops of fictitious fermionic particles, the Faddeev-Popov
ghosts \cite{fadpop}. The key role of DeWitt in this context 
was emphasized by Veltman in 1974 \cite{veltman}:
\begin{quote}
    \ldots Essentially due to this, and some deficiencies in his 
    combinatorial methods, Feynman was not able to go beyond one closed 
    loop.  DeWitt in his 1964 Letter and in his subsequent monumental 
    work derived most of the things that we know of now. That is, he 
    consider the question of a choice of gauge and the associated 
    ghost particle. Indeed, he writes the ghost contribution in the 
    form of a local Lagrangian containing a complex scalar field 
    obeying Fermi statistics. Somewhat illogically this ghost is now 
    called the Faddeev-Popov ghost. 
\end{quote} 
On the other hand, however, in comparison with the complicated
combinatorics of DeWitt, the Faddeev-Popov approach has the merit of a
greater technical simplicity and of a transparent geometrical
interpretation, which are probably the reason for its popularity.  It
is in the work of Faddeev that the key role played by the gauge orbits
(and not fields at a given point) as true dynamical variables, is
fully elucidated \cite{spires}.

\vskip2mm{\bf 1967} \nopagebreak

Bryce DeWitt publishes the ``Einstein-Schr\"odinger equation"
\cite{dewitt67}.  
$$
\left((\hbar G)^{2}(q_{ab}q_{cd}-{1\over 2}q_{ac}q_{bd}){\delta \over
\delta q_{ac}} {\delta \over\delta q_{bd}}-\det q\, R[q]\right)\
\Psi(q)=0,
$$
Bryce will long denote this equation as the ``Einstein-Schr\"odinger
equation", attributing it to Wheeler --while John Wheeler denoted it
as the DeWitt equation-- until, finally, in 1988, at a Osgood Hill
conference, DeWitt gives up and calls it the way everybody else had
been calling it since the beginning: the ``Wheeler-DeWitt equation".

The story of the birth of the Wheeler-DeWitt equation is worth
telling.  In 1965, during an air trip, John had to stop for a short
time at the Rahley-Durham airport in Noth Carolina.  Bryce lived
nearby.  John phoned Bryce and proposed to meet at the airport during
the wait between two planes.  Bryce showed up with the Hamilton-Jacobi
equation for GR, published by Peres in 1962 and mumbled the idea of
doing precisely what Shr\"odinger did for the hydrogen atom: replace
the square of the derivative with a second derivative.  Surprising
Bryce, John was enthusiastic (John is often enthusiastic, of course),
and declared immediately that {\em the\/} equation of quantum gravity
had been found.  The paper with the equation, the first of Bryce's
celebrated 1967 quantum gravity trilogy \cite{dewitt67,trilogy}, was
submitted in the spring of 66, but its publication was delayed until
1967.  Apparently, also because of difficulties with publication
charges \ldots

\vskip2mm{\bf 1967} \nopagebreak

John Wheeler discusses the idea of wave function $\Psi(q)$ of the 
``3-geometry" $q$, and the notion of superspace, the space of the 
3-geometries in \cite{wheeler67}.

Penrose starts twistor theory \cite{twistors}. 

The project of DeWitt and Feynman is concluded.  A complete and
consistent set of Feynman rules for GR are written down
\cite{trilogy,fadpop}.

\vskip2mm{\bf 1969} \nopagebreak 

Developing an idea in Bryce's paper on canonical quantum gravity,
Charles Misner starts quantum cosmology: the game of truncating the
Wheeler-DeWitt equation to a finite number of degrees of freedom
\cite{qcosmology}.  The idea is beautiful, but it will develop into a
long lasting industry from which, after a while, little new will be
understood.

\vskip1cm

The decade closes with the main lines of the covariant and the canonical 
theory clearly defined. It will soon become clear that neither theory 
works. 

\section{The Middle Ages: 1970-1983}

\hspace{.6cm}{\bf 1970} \nopagebreak

The decade of the seventies opens with a world of caution.  
Reviving a point made by Pauli, a paper by Zumino \cite{zumino}, 
suggests that the quantization of GR may be problematic and might 
make sense only by viewing GR as the low energy limit of a more 
general theory.

\vskip2mm{\bf 1971} \nopagebreak

Using the technology developed by DeWitt and Feynman for gravity, 
t'Hooft and Veltman decide to study the renormalizability of GR. 
Almost as a warm up exercise, they consider the renormalization 
of Yang-Mills theory, and find that the theory is renormalizable 
-- result that has won them this year Nobel prize 
\cite{t'Hooft71}.  In a sense, one can say that the first 
physical result of the research in quantum gravity is the proof 
that Yang-Mills theory is renormalizable.

\vskip2mm{\bf 1971} \nopagebreak

David Finkelstein writes his inspiring ``spacetime code" series of
papers \cite{Finkelstein} (which, among others ideas, discuss quantum
groups).

\vskip2mm{\bf 1973} \nopagebreak

Following the program, t'Hooft finds evidence of un-renormalizable 
divergences in GR with matter fields.  Shortly after, t'Hooft and 
Veltman, as well as Deser and Van Nieuwenhuizen, confirm the 
evidence \cite{thooft73}.

\vskip2mm{\bf 1974} \nopagebreak

Hawking announces the derivation of black hole radiation 
\cite{hawking74}.  A (macroscopically) Schwarzshild black hole of 
mass $M$ emits thermal radiation at the temperature
$$
       T = {\hbar c^{3}\over 8\pi k G M} 
$$
The result comes as a surprise, anticipated only by the 
observation by Bekenstein, a year earlier, that entropy is 
naturally associated to black holes, and thus they could be 
thought, in some obscure sense, as ``hot" \cite{bekenstein}, and 
by the Bardeen-Carter-Hawking analysis of the analogy between 
laws of thermodynamics and dynamical behavior of black holes.  
Hawking's result is not directly connected to quantum gravity 
--it is a skillful application of quantum field theory in curved 
spacetime-- but has a very strong impact on the field.  It 
fosters an intense activity in quantum field theory in curved 
spacetime, it opens a new field of research in ``black hole 
thermodynamics" (for a review of the two, see \cite{wald}), and 
it opens the quantum-gravitational problems of understanding the 
statistical origin of the black hole (the Bekenstein-Hawking) 
entropy. For a Schwarzshild black hole, this is  
\begin{equation}
       S = {1\over 4}\, {c^{3}  \over \hbar G}\, A 
\label{entropy}
\end{equation}
where $A$ is the area of the black hole surface.  An influential,
clarifying and at the same time intriguing paper is written two years
later by Bill Unruh.  The paper points out the existence of a general
relation between accelerated observers, quantum theory, gravity and
thermodynamics \cite{unruh}.  Something deep about nature should be
hidden in this tangle of problems, but we do not yet know what.

\vskip2mm{\bf 1975} \nopagebreak

It becomes generally accepted that GR coupled to matter is not
renormalizable.  The research program started with Rosenfeld, Fierz
and Pauli is dead.

\vskip2mm{\bf 1976} \nopagebreak

A first attempt to save the covariant program is made by Steven 
Weinberg, who explore the idea of asymptotic safety 
\cite{weinberg76}, developing earlier ideas from Giorgio Parisi 
\cite{parisi}, Kenneth Wilson and others, suggesting that 
non-renormalizable theories could nevertheless be meaningful.

\vskip2mm{\bf 1976} \nopagebreak

To resuscitate the covariant theory, even if in modified form, 
the path has already been indicated: find a high energy 
modification of GR. Preserving general covariance, there is not 
much one can do to modify GR. An idea that attracts much 
enthusiasm is supergravity \cite{supergravity}: it seems that by 
simply coupling a spin 3/2 particle to GR, namely with the action 
(in first order form)
$$
 S[g,\Gamma,\psi] = \int d^{4}x\ \sqrt{-g}\ \left({1\over 2G} R - 
 {i\over 2} \ \epsilon^{\mu\nu\rho\sigma}\ \psi_{\mu} \gamma_{5} 
 \gamma_{\nu} D_{\rho} \psi_{\sigma} \right),
$$
one can get a theory finite even at two loops. 

\vskip2mm{\bf 1977} \nopagebreak

Another, independent, idea is to keep the same kinematics and change
the action.  The obvious thing to do is to add terms proportional to
the divergences.  Stelle proves that an action with terms quadratic in
the curvature
$$
 S = 
 \int d^{4}x\ \sqrt{-g}\ 
 \left(\alpha R+\beta R^{2}+\gamma R^{\mu\nu}R_{\mu\nu} . 
 \right),
 $$
is renormalizable for appropriate values of the coupling constants
\cite{stelle}.  Unfortunately, precisely for these values of the
constants the theory is bad.  It has negative energy modes that make
it unstable around the Minkowski vacuum and not unitary in the quantum
regime.  The problem becomes to find a theory renormalizable and
unitary at the same time, or to circumvent non-unitarity. 

\vskip2mm{\bf 1978} \nopagebreak

The Hawking radiation is soon re-derived in a number of ways, 
strongly reinforcing its credibility.  Several of these 
derivations point to thermal techniques \cite{hht}, thus 
motivating Hawking \cite{haw} to revive the Wheeler-Misner 
``Feynman quantization of general relativity" \cite{misner} in 
the form of a ``Euclidean" integral over {\em Riemannian\/} 
4-geometries $g$
$$
  Z = \int \ Dg\ e^{-\int \sqrt{g}R}. 
$$
Time ordering and the concept of positive frequency are 
incorporated into the ``analytic continuation" to the Euclidean 
sector.  The hope is double: to deal with topology change, and 
that the Euclidean functional integral will prove to be a better 
calculation tool than the Wheeler-DeWitt equation.

\vskip2mm{\bf 1980} \nopagebreak

Within the canonical approach, the discussion focuses on understanding
the disappearance of the time coordinate from the Wheeler-DeWitt
theory.  The problem has actually nothing to do with {\em quantum\/}
gravity, since the time coordinate disappears in the {\em classical\/}
Hamilton-Jacobi form of GR as well; and, in any case, physical
observables are coordinate independent, and thus, in particular,
independent from the time coordinate, in whatever correct formulation
of GR. But in the quantum context there is no single spacetime, as
there is no trajectory for a quantum particle, and the very concepts
of space and time become fuzzy.  This fact raises much confusion and a
vast interesting discussion (whose many contributions I can not
possibly summarize here) on the possibility of doing meaningful
fundamental physics in the absence of a fundamental notion of time. 
For early references on the subject see for instance \cite{time}.

\vskip2mm{\bf 1981} \nopagebreak

Polyakov \cite{polyakov} shows that the cancellation of the conformal
anomaly in the quantization of the string action
$$
    S = {1\over 4\pi\alpha'} \int d^{2}\sigma \sqrt{g} 
    \ g^{\mu\nu} 
    \partial_{\mu}X^{a} \partial_{\nu}X^{b}\eta_{ab}.
$$
leads to the critical dimension.  

\vskip2mm{\bf 1983} \nopagebreak

The hope is still high for supergravity, now existing in various
versions, as well as for higher derivative theories, whose rescue from
non-unitarity is explored using a number of ingenious ideas (large N
expansions, large D expansions, Lee-Wick mechanisms\ldots).  At the
10th GRG conference in Padova in 1983, two physicists of indisputable
seriousness, Gary Horowitz and Andy Strominger, summarize their
contributed paper \cite{horowitzstrominger} with the words
\begin{quote} 
    In sum, higher derivative gravity theories are a viable option for 
    resolving the problem of quantum gravity \ldots 
\end{quote} 
At the same conference, supergravity is advertised as a likely 
final solution of the quantum gravity puzzle.  But very soon it 
becomes clear that supergravity is non-renormalizable at higher 
loops and that higher derivatives theories do not lead to viable 
perturbative expansions.  The excitement and the hope fade away.  

In its version in 11 dimensions, supergravity will find new 
importance in the late 1990s, in connection with string theory.  
High derivative corrections will also reappear, in the low energy 
limit of string theory.

\vskip2mm{\bf 1983} \nopagebreak

Hartle and Hawking \cite{hartlehawking} introduce the notion of 
the ``wave function of the universe" and the ``no-boundary" 
boundary condition for the Hawking integral, opening up a new 
intuition on quantum gravity and quantum cosmology.  But the 
Euclidean integral does not provide a way of computing genuine 
field theoretical quantities in quantum gravity better than the 
Wheeler-DeWitt equation, and the atmosphere at the middle of the 
eighties is again rather gloomy.  On the other hand, Jim Hartle 
\cite{hartle} develops the idea of a sum over histories 
formulation of GR into a full fledged extension of quantum 
mechanics to the general covariant setting.  The idea will later 
be developed and formalized by Chris Isham \cite{Isham}.

Sorkin introduces his poset approach to quantum gravity \cite{sorkin}.

\section{The Renaissance: 1984-1994}

\hspace{.6cm}{\bf 1984} \nopagebreak

Green and Schwarz realize that strings might describe ``our universe"
\cite{green}.  Excitement starts to build up around string theory, in
connection with the unexpected anomaly cancellation and the discovery
of the heterotic string \cite{hetero}.

The relation between the ten dimensional superstrings theory and four
dimensional low energy physics is studied in terms of compactification
on Calaby-Yau manifolds \cite{calabi} and orbifolds.  The dynamics of
the choice of the vacuum remains unclear, but the compactification
leads to 4d chiral models resembling low energy physics.

Belavin, Polyakov and Zamolodchikov publish their analysis of 
conformal field theory \cite{conformal}.

\vskip2mm{\bf 1986}  \nopagebreak

Goroff an Sagnotti \cite{sagnotti} finally compute the two loop
divergences of pure GR, definitely nailing the corpse of pure GR
perturbative quantum field theory into its coffer: the divergent term 
is
$$
 \Delta S = {209\over 737280\pi^{4}}\  {1\over\epsilon}
 \int d^{4}x\ \sqrt{-g}\ 
 R^{\mu\nu}{}_{\rho\sigma}R^{\rho\sigma}{}_{\epsilon\theta} 
 R^{\epsilon\theta}{}_{\mu\nu} . 
$$

\vskip2mm{\bf 1986}  \nopagebreak

Penrose suggests that the wave function collapse in quantum mechanics
might be of quantum gravitational origin \cite{penrose}.  The idea is
radical and implies a re-thinking of the basis of mechanics. 
Remarkably, the idea may be testable: work is today in progress to
study the feasibility of an experimental test.

\vskip2mm{\bf 1986}  \nopagebreak

String field theory represents a genuine attempt to address the main
problem of string theory: finding a fundamental, background
independent, definition of the theory \cite{stringfield}.  The string
field path, however, turns out to be hard.

\vskip2mm{\bf 1986}  \nopagebreak

The connection formulation of GR is developed by Abhay Ashtekar 
\cite{ashtekar}, on the basis of some results by Amitaba Sen 
\cite{sen}.  At the time, this is denoted the ``new variables" 
formulation.  It is a development in classical general 
relativity, but it has long ranging consequences on quantum 
gravity, as the basis of loop quantum gravity.

\vskip2mm{\bf 1987}  \nopagebreak

Fredenhagen and Haag explore the general constraint that general 
covariance puts on quantum field theory \cite{haag}. 

\vskip2mm{\bf 1987}  \nopagebreak

Green, Schwarz and Witten publish their book on superstring 
theory.  In the gauge in which the metric has no superpartner, 
the superstring action is
$$ 
    S = {1\over 4\pi\alpha'} \int d^{2}\sigma \sqrt{g} 
    \left( g^{\mu\nu} 
    \partial_{\mu}X^{a} \partial_{\nu}X^{b} - i \psi^a
    \gamma^{\mu}\partial_{\mu}\psi^b\right) \eta_{ab}.
$$
The interest in the theory grows very rapidly.  To be sure, 
string theory still obtains a very small place at the 1991 Marcel 
Grossmann meeting \cite{MG91}.  But, increasingly, the research 
in supergravity and higher derivative theories has merged into 
strings, and string theory is increasingly viewed as a strong 
competing candidate for the quantum theory of the gravitational 
field.  As a side product, many particle physicists begin to 
study general relativity, or at least some bits of it.  Strings 
provide a consistent perturbative theory.  The covariant program 
is fully re-born.  The problem becomes understanding why the 
world described by the theory appears so different from ours.

\vskip2mm{\bf 1988}  \nopagebreak

Ted Jacobson and Lee Smolin find loop-like solutions to the
Wheeler-DeWitt equation formulated in the connection formulation
\cite{TL}, opening the way to loop quantum gravity. 

\vskip2mm{\bf 1988}  \nopagebreak

The ``loop representation of quantum general relativity" is introduced
in \cite{rovellismolin88}.  For an early review, see \cite{report}. 
It is based on the new connection formulation of GR \cite{ashtekar},
on the Jacobson-Smolin solutions \cite{TL}, and on Chris Isham's ideas
on the need of non-Gaussian, or non-Fock representations in quantum
gravity \cite{chris}.  Loop quantization had been previously and
independently developed by Rodolfo Gambini and his collaborators for
Yang Mills theories \cite{gambini}.  In the gravitational context, the
loop representation leads immediately to two surprising results: an
infinite family of exact solutions of the Wheeler-DeWitt equation is
found, and knot theory controls the physical quantum states of the
gravitational field.  Classical knot theory, with its extensions,
becomes a branch of mathematics relevant to describe the diff
invariant states of quantum spacetime \cite{pullin}.

The theory transforms the old Wheeler-DeWitt theory in a formalism
that can be concretely used to compute physical quantities in quantum
gravity.  The canonical program is fully re-born.  Today, the theory
is called ``loop quantum gravity".\footnote{It is sometimes called
also ``quantum geometry", but the expression ``quantum geometry" is
used by a variety of other research programs as well.} For a review,
complete references, and an account of the development of the theory,
see \cite{lqg}.

\vskip2mm{\bf 1988}  \nopagebreak

Ed Witten introduces the notion of topological quantum field theory
(TQFT) \cite{wittentqft}.  In a celebrated paper \cite{wittentjones},
he uses a TQFT to give a field theoretical representation of the Jones
polynomial, a knot theory invariant.  The expression used by Witten
has an interpretation in loop quantum gravity: it can be seen as the
``loop transform" of quantum state given by the exponential of the
Chern Simon functional \cite{pullin}. 

Formalized by Atiyah \cite{atiyahtqft}, the idea of TQFT will have
beautiful developments, and will strongly affect later development in
quantum gravity.  General topological theories in any dimensions, and
in particular BF theory, are introduced by Gary Horowitz shortly
afterwards \cite{garyBF}.

\vskip2mm{\bf 1988}  \nopagebreak

Witten finds an ingenious way of quantizing GR in 2+1 spacetime
dimensions \cite{witten2+1}, (thus solving ``problem 56" of the 1963
Wheeler's Les Houches lectures) opening up a big industry of analysis
of the theory (for a review, see \cite{carlip}).  The quantization
method is partially a sum over histories and partially canonical. 
Covariant perturbative quantization seemed to fail for this theory. 
The theory had been studied a few years earlier by Deser, Jackiw, 
t'Hooft, Achucarro, Townsend and others \cite{dj}.

\vskip2mm{\bf 1989}  \nopagebreak

Amati, Ciafaloni and Veneziano find evidence that string theory implies
that distances smaller than the Planck scale cannot be probed
\cite{amativeneziano}. 

\vskip2mm{\bf 1989}  \nopagebreak

In the string world, there is  excitement for some
nonperturbative models of strings ``in 0 dimension", equivalent to 2D
quantum gravity \cite{2d}.  The excitement dies fast, as many others,
but the models will re-emerge in the nineties \cite{2D2}, and will also
inspire the spin foam formulation of quantum gravity \cite{spinfoams}.

\vskip2mm{\bf 1992}  \nopagebreak

Turaev and Viro \cite{turaevviro} define a state sum that on the one
hand is a rigorously defined TQFT, on the other hand can be seen as a
regulated and well defined version of the Ponzano-Regge
\cite{Ponzano:1968} quantization of 2+1 gravity.  Turaev, and Ooguri
\cite{Ooguri:1992b} find soon a 4d extension, which will have a
remarkable impact on later developments.

\vskip2mm{\bf 1992}  \nopagebreak

The notion of {\em weave\/} is introduced in loop quantum gravity
\cite{weave}.  It is evidence of a discrete structure of spacetime
emerging from loop quantum gravity.  The first example of a weave
which is considered is a 3d mesh of intertwined rings.  Not
surprising, the intuition was already in Wheeler!  See Figure 1, taken
from Misner Thorne and Wheeler \cite{mtw}.

\begin{figure}[h]
\centerline{{\psfig{figure=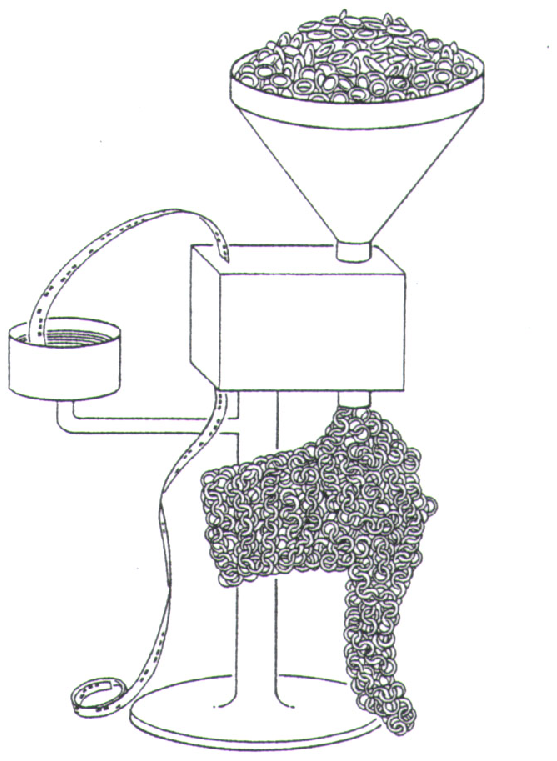,height=11cm}}}
\caption{The weave, in Wheeler's vision.}
\end{figure}

\vskip2mm{\bf 1994}  \nopagebreak

Noncommutative geometry, often indicated as a tool for describing
certain aspects of Planck scale geometry, finds a strict connection to
GR in the framework of Alain Connes' noncommutative geometry. 
Remarkably, the Connes-Chamseddine ``spectral action", just the trace
of a simple function of a suitably defined Dirac-like operator $D$
$$
  S = Tr[f(D^{2}/(\hbar G))],
$$
where $f$ is the characteristic function of the $[0,1]$ interval, 
turns out to include the standard model action, as well as the 
Einstein-Hilbert action \cite{conneschamseddine}.

\section{Nowadays: 1995-2000}

\hspace{.6cm}{\bf 1995}  \nopagebreak

Nonperturbative aspects of string theory begin to appear: branes
\cite{polchi}, dualities \cite{hull}, the matrix model
formulation of M theory \cite{m} \ldots .  (For a review, see for
instance \cite{duff}).  The interest in strings booms.  At the plenary
conference of a meeting of the American Mathematical Society in
Baltimore, Ed Witten claims that
\begin{quote}
    ``The mathematics of the next millennium will be dominated 
    by string theory"
\end{quote}
causing a few eyebrows to raise.

The various dualities appear to relate the different versions of the
theory, pointing to the existence of a unique fundamental theory.  The
actual construction of the fundamental background independent theory,
however, is still missing, and string theory exists so far only in the
form of a number of (related) expansions over assigned backgrounds.

\vskip2mm{\bf 1995}  \nopagebreak

Using the spin network orthonormal basis found on the Hilbert 
space of loop quantum gravity, a main physical result is obtained 
within loop quantum gravity: the computation of the eigenvalues 
of area and volume \cite{rovellismolin95b}.  The main sequence of 
the eigenvalues of the area is labeled by an $n$-tuplet of half 
integers $\vec j=\{j_{1}\ldots j_{n}\}$ and is
$$
    A_{\vec j} = 8\pi\hbar G\ \sum_{i=1,n}\sqrt{j_{i}(j_{i}+1)}. 
$$
The result is rapidly extended and derived in a number of ways
\cite{area}. 

A rigorous mathematical framework for loop quantum gravity is
developed \cite{md}.

\vskip2mm{\bf 1995} \nopagebreak
 
Ted Newman and his collaborators introduce the Null Surface 
Formulation of GR \cite{Newman}. 

\vskip2mm{\bf 1996}  \nopagebreak

The Bekenstein-Hawking black hole entropy is computed within loop
quantum gravity as well as within string theory, almost at the same
time.  

The loop result is obtained by computing the number of (spin-network)
states which endow a 2-sphere with a given area \cite{loopbh}, as well
as by loop quantizing the classical theory of the field outside the
hole and studying the boundary states \cite{loopbh2}.  These
gravitational surface states \cite{surface} can be identified with the
states of a Chern-Simons theory on a surface with punctures
\cite{leebh}.  The computation is valid for various realistic black
holes, and the 1/4 factor in (\ref{entropy}) is obtained by fixing the
undetermined dimensionless parameter present in loop quantum gravity
(the Immirzi parameter).

In string theory, the computation exploits a strong coupling/weak
coupling duality, which, in certain supersymmetric configurations,
preserves the number of states: the physical black hole is in a strong
coupling situation, but the number of its microstates can be computed
in a weak field configuration that has the same charges at infinity. 
The calculation method is thus rather indirect, and works smoothly
only for certain extremal black holes; remarkably, however, one
obtains precisely the 1/4 factor of equation (\ref{entropy}), as well
as other key aspects of the Hawking radiation phenomenology
\cite{stromingervafa}.

\vskip2mm{\bf 1996}  \nopagebreak

A rigorously defined, finite and anomaly free hamiltonian constraint
operator is constructed by Thomas Thiemann in loop quantum gravity
\cite{thiemann}.  Some doubts are raised on whether the classical
limit of this theory is in fact GR (the issue is still open), but the
construction defines a consistent general covariant quantum field
theory in 4d.

\vskip2mm{\bf 1996}  \nopagebreak

Intriguing state sum models obtained modifying a TQFT are proposed by
Barrett and Crane, Reisenberger, Iwasaki and others \cite{spinfoams}
as a tentative model for quantum GR. All these models appear as sums
of ``spin foams": branched surfaces carrying spins.  

In the meanwhile, the loop representation is ``exponentiated", \`a la
Feynman, giving rise, again, to a spin foam model, corresponding to
canonical loop quantum gravity \cite{reisenbergrovelli}.  These
developments revive the sum over histories approach.

\vskip2mm{\bf 1997}  \nopagebreak

There is a lively discussion on the difficulties of the lattice
approaches in finding a second order phase transition \cite{Carfora}.

Intriguing connections between non commutative geometry and string 
theory appear \cite{stringnoncomm}.

\vskip2mm{\bf 1998} \nopagebreak

Juan Maldacena shows \cite{maldacena} that the large $N$ limit of
certain conformal field theories includes a sector describing
supergravity on the product of Anti-deSitter spacetimes and spheres. 
He conjectures that the compactifications of M/string theory on an
Anti-deSitter spacetimes is dual to a conformal field theory on the
spacetime boundary.  This leads to a new proposal for defining
M-theory itself in term of the boundary theory: an effort to reach
background independence (for M theory) using background dependent
methods (for the boundary theory).

A consequence of this ``Maldacena conjecture" is an explosion of
interest for an idea by Gerard t'Hooft, developed and promoted by
Leonard Suskind: the ``holographic principle".  According to this
principle (considered in a number of variants) the information on the
physical state in the interior of a region can be represented on the
region's boundary and is limited by the area of this boundary.

\vskip2mm{\bf 1998} \nopagebreak

Two papers in the influential journal {\em Nature\/}
\cite{amelinocamelia} raise the hope that seeing spacetime-foam
effects, and testing quantum gravity theories might not be as
forbidding as usually assumed.  The idea is that there is a number of
different instances (the neutral kaon system, gamma ray burst
phenomenology, interferometers \ldots) in which presently operating
measurement or observation apparata, or apparata that are going to be
soon constructed, involve sensitivity scales comparable to --or not
too far from-- the Planck scale \cite{phenomenology}.  If this 
direction fails, testing quantum gravity might require the 
investigation of very early cosmology \cite{veneziano}. 

\vskip2mm{\bf Today }\nopagebreak

For critical discussions of current direction of research in quantum
gravity, see for instance \cite{qg}.

\section{Concluding remarks}

The lines of research that I have summarized in Section 2 have
found many points of contact in the course of their development and
have often intersected each other.  For instance, there is a formal
way of deriving a sum over over histories formulation from a canonical
theory and viceversa; the perturbative expansion can also be obtained
expanding the sum over histories; string theory today faces the
problem of a finding its nonperturbative formulation, and thus the
typical problems of a canonical theory, while loop quantum gravity has
mutated into the spin foam models, a sum over history formulation,
using techniques that can be traced to a development of string theory
of the early nineties.  Recently, Lee Smolin has been developing an
attempt to connect nonperturbative string theory and loop quantum
gravity \cite{leemerge}.  However, in spite of this continuous cross
fertilization, the three main lines of development have kept their
essential separation.

As pointed out, the three direction of investigation where 
already clearly identified by Charles Misner in 1959 
\cite{misner}.  In the concluding remark of the {\em Conf\'erence 
internationale sur les th\'eories relativistes de la 
gravitation}, in 1963, Peter Bergmann noted \cite{pologne}
\begin{quote}
  ``In view of the great difficulties of this program, I consider it 
    a very positive thing that so many different approaches are being 
    brought to bear on the problem.  To be sure, the approaches, we 
    hope, will converge to one goal."  
\end{quote}
This was almost 40 years ago \ldots
 
The divide is particularly macroscopic between the covariant line
of research on the one hand and the canonical/sum over histories on
the other.  This divide has remained through over 70 years of research
in quantum gravity.  The separation cannot be stronger.  Here is a
typical comparison, arbitrarily chosen among many.  On the covariant
side, at the First Marcel Grossmann Meeting, Peter van Neuwenhuizen
writes \cite{vnw}
\begin{quote}
    ``\ldots gravitons are treated on exactly the same basis as other
    particles such as photons and electrons.  In particular, particles
    (including gravitons) are always in flat Minkowski space and move
    {\em as if\/} they followed their geodesics in curved spacetime
    because of the dynamics of multiple graviton exchange.  [\ldots]
    Pure relativists often become somewhat uneasy at this point because
    of the the following two aspects entirely peculiar to gravitation:
    1) [\ldots].  One must decide before quantization which points are
    spacelike separated, but  it is only after
    quantization that that the fully quantized metric field can tell
    us this spacetime structure [\ldots].  2) [\ldots] In a classical
    curved background one needs positive and negative solutions, but in
    non-stationary spacetimes it is not clear whether one
    can define such solutions.  {\em The strategy of particle
    physicists has been to ignore these problems for the time being,
    in the hope that they will ultimately be resolved in the final
    theory.  Consequently we will not discuss them any further.}"
\end{quote}
On the canonical side\footnote{For a detailed defense of the
relativist point of view in the debate, see \cite{me}.}, Peter
Bergmann \cite{pb}
\begin{quote}
``The world point by itself possess no physical reality. It acquires 
reality only to the extent that it becomes the bearer of specific 
properties of the physical fields imposed on the spacetime manifold."  
\end{quote}
Partially, the divide reflects the different understanding of the
world that the particle physics community on the one hand and the
relativity community on the other hand, have.  The two communities
have made repeated and sincere efforts to talk to each other and
understanding each other.  But the divide remains, and, with the
divide, the feeling, on both sides, that the other side is incapable
of appreciating something basic and essential: On the one side, the
structure of quantum field theory as it has been understood in half a
century of investigation; on the other side, the novel physical
understanding of space and time that has appeared with general
relativity.  Both sides expect that the point of the other will turn
out, at the end of the day, to be not very relevant.  One side because
all the experience with quantum field theory is on a fixed metric
spacetime, and thus is irrelevant in a genuinely background
independent context.  The other because GR is only a low energy limit
of a much more complex theory, and thus cannot be taken too seriously
as an indication on the deep structure of Nature.  Hopefully, the
recent successes of both lines will force the two sides, finally, to
face the problems that the other side considers prioritary: background
independence on the one hand, control of a perturbation expansion on
the other. 

\vskip.5cm

So, where are we, after 70 years of research?  There are
well-developed tentative theories, in particular strings and loops,
and several other intriguing ideas.  There is no consensus, no
established theory, and no theory that has yet received any direct or
indirect experimental support.  In the course of 70 years, many ideas
have been explored, fashions have come and gone, the discovery of the
Holly Graal has been several times announced, with much later scorn.

However, in spite of its age, the research in quantum gravity 
does not seem to have been meandering meaninglessly, when seen in 
its entirety.  On the contrary, one sees a logic that has guided 
the development of the research, from the early formulation of 
the problem and the research directions in the fifties to 
nowadays.  The implementation of the programs has been extremely 
laborious, but has been achieved.  Difficulties have appeared, 
and solutions have been proposed, which, after much difficulty, 
have lead to the realization, at least partial, of the initial 
hopes.  It was suggested in the early seventies that GR could 
perhaps be seen as the low energy theory of a theory without 
uncontrollable divergences; today, 30 years later, such a theory 
--string theory-- is known.  In 1957 Charles Misner indicated 
that in the canonical framework one should be able to compute 
eigenvalues; and in 1995, 37 years later, eigenvalues were 
computes --within loop quantum gravity.  The road is not yet at 
the end, much remains to be understood, and some of the current 
developments might lead nowhere.  But it is difficult to deny, 
looking at the entire development of the subject, that there has 
been a linear progress.  And the road, no doubts, is fascinating.

\vskip1cm

I am very much indebted to many friends that have contributed to this
reconstruction.  I am particularly grateful to Augusto Sagnotti, Gary
Horowitz, Ludwig Faddeev, Alejandro Corichi, Jorge Pullin, Lee Smolin,
Joy Christian, Bryce DeWitt, Giovanni Amelino-Camelia, Daniel
Grumiller, Nikolaos Mavromatos and Ted Newman.

\end{document}